%% file: main.tex
\documentclass[conference]{IEEEtran}

\usepackage{cite, algorithmic, amsfonts, amssymb, amsmath, float, graphicx, url, hyperref, listings, subfig, mathtools, color, multirow, multicol, textcomp, enumerate, amsthm, cancel}

\def\BibTeX{{\rm B\kern-.05em{\sc i\kern-.025em b}\kern-.08em
    T\kern-.1667em\lower.7ex\hbox{E}\kern-.125emX}}

\DeclareMathOperator*{\group}{gr}
\DeclareMathOperator*{\leaf}{leaf}
\DeclareMathOperator*{\totalhosts}{N}
\DeclareMathOperator*{\amt}{rate}

\begin{document}

\newcommand{\bigparen}[1]{\Bigl ( #1 \Bigr )}
\newcommand{\bigparensq}[1]{\Bigl [ #1 \Bigr ]}

\newcommand{\minisep}[0]{\vspace{-0.15cm}\rule{2cm}{0.4pt}\vspace{0.05cm}}
\newcommand{\RNum}[1]{\uppercase\expandafter{\romannumeral #1\relax}}
\newcommand{\eqalign}[1]{\vspace{-0.5cm} \begin{equation*} \begin{aligned} #1 \end{aligned} \phantom{\hspace{15cm}} \end{equation*}}

\newcommand\applypropone{\stackrel{\mathclap{\tiny\mbox{\bf p. IV.1}}}{\leq}}

\theoremstyle{definition}
\newtheorem{statement}{Statement}[section]
\newtheorem{lemma}[statement]{Lemma}
\newtheorem{property}[statement]{Property}
\newtheorem{notation}[statement]{Notation}
\newtheorem{theorem}[statement]{Theorem}
\newtheorem{assumption}[statement]{Assumption}
\newtheorem{observation}[statement]{Observation}
\newtheorem{overview}[statement]{Idea Overview}
\newtheorem{explanation}[statement]{Explanation}
\newtheorem{section_overview}[statement]{Section Overview}
\newtheorem{mistake}[statement]{Mistake}
\newtheorem{question}[statement]{Question}

\lstset{frame=tb,
  language=Python,
  aboveskip=3mm,
  belowskip=3mm,
  showstringspaces=false,
  columns=flexible,
  basicstyle={\small\ttfamily},
  numbers=none,
  numberstyle=\tiny\color{gray},
  keywordstyle=\color{blue},
  commentstyle=\color{magenta},
  stringstyle=\color{red},
  breaklines=true,
  breakatwhitespace=true,
  tabsize=3
}

\title {Understanding the oversubscription behaviour of DragonFly+ networks}

\author{
  \IEEEauthorblockN{Vlad-Adrian Ulmeanu$^1$, Costin Raiciu$^{1, 2}$, and Iulian-Ilie Drăcea$^2$}
  \IEEEauthorblockA{
    University Politehnica of Bucharest$^1$, Broadcom$^2$\\
    \texttt{\{\href{mailto:vlad_adrian.ulmeanu@stud.acs.upb.ro}{vlad\_adrian.ulmeanu@stud.acs.}}, \texttt{\href{mailto:costin.raiciu@upb.ro}{costin.raiciu@}\}upb.ro}, \texttt{\href{mailto:iulian-ilie.dracea@broadcom.com}{iulian-ilie.dracea@broadcom.com}}
  }
}

\maketitle

\begin{abstract}

The Max-Host Dragonfly+ topology's original paper proves that there is a 2:1 worst-case oversubscription ratio in expectation for the permutation traffic pattern. We show that the proof only covers a subset of permutation patterns, specifically those in which all host pairs are in different groups, and for any receiver group there are at least two sending hosts in distinct groups. The proof contains two mistakes that cancel out to produce the correct result. Furthermore, the proof implicitly uses an important observation without a backing argument: traffic leaving an indirect group, now forced to follow min-cost paths to receiver groups, can still be split across almost all outgoing global links, in contrast with splitting traffic from a sender group that can naturally select any outgoing group.

We prove that the 2:1 ratio holds in expectation for a larger set of patterns, including any permutation. We only need to know that any host sends and receives at most line rate traffic. This constraint can be altered to obtain approximations for expected FCT bounds on any pattern. We also attack the oversubscription problem without the expectation assumption, and find bounds that hold with high probability on permutations for small switch radixes.

While in expectation the global layer wiring doesn't affect the oversubscription rate, we experimentally find that it matters in the general case. We find topologies that obtain visible speedups against the default global wiring from the original paper, under $7\%$ for radixes at most $8$.

\end{abstract}

\input{tex_parts/introduction}

\input{tex_parts/problem_definition}

\input{tex_parts/dfp_proof_discussion}

\input{tex_parts/2_1_oversub}

\input{tex_parts/2_1_oversub_high_proba_bound}

\input{tex_parts/conclusion}

\section*{Acknowledgment}

We thank the anonymous reviewers for their constructive feedback. We thank our faculty for providing the compute cluster used for simulations.

\newpage

\bibliographystyle{plainurl}
\bibliography{bibliography}

\end{document}

%% file: tex_parts/introduction.tex
\section{Introduction}\label{section_introduction}

Low-diameter datacenter topologies that either accommodate more hosts, or increase the number of redundant paths between two nodes for a fixed switch radix are preferred for LLM training workloads, which can currently make use of tens of thousands of GPUs if training from scratch. The most well-known low-diameter topology is the Dragonfly \cite{Dragonfly}. Currently 6 of the top 10 supercomputers in the top 500 list \cite{Top500} use a derivative of it (Cray Slingshot-11). Nvidia is expected to adopt the Dragonfly topology for the next-generation scale-up networks (NVL288 and NVL576) \cite{NvidiaMayUseDfly}, trading a higher oversubscription ratio for keeping the same switch radix, in contrast with the single-tier multi-plane topology currently found in e.g. NVL72.

\begin{figure}[H]
    \centering
    \includegraphics[width=0.4\linewidth]{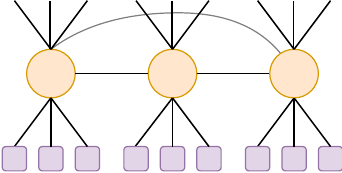}
    \caption{Nodes are organized in groups in the Dragonfly topology.}
    \label{introDflyGroup}
\end{figure}

A group has two levels: the first one is made out of hosts. Each host is connected to a switch on the secondary level. These switches form a clique or mesh. The switches' outgoing pipes unite the group with other ones. There is at least one link (named global) between any pair of groups.\\

Notice that Dragonfly can create many more paths between two specific hosts by not selecting the min-cost switch in the sender group. As a result, a packet may arrive in an intermediate group, from which it will take the min-cost path to the destination.

\begin{figure}[H]
    \centering
    \includegraphics[width=0.4\linewidth]{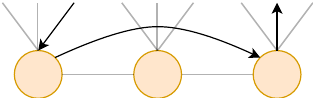}
    \caption{The packet might (or not) need to change the outgoing switch, depending on the topology wiring. Getting to the correct switch takes one extra hop because of the clique structure. We distinguish between routes that don't need to change the switch and the ones that do by calling the first 1-indirect and the second 2-indirect.}
    \label{introDflyGroupSwitch}
\end{figure}

Some well-known offsprings of the Dragonfly are the Slimfly \cite{Slimfly}, \cite{EdgeDisjointSpanningTrees} and Dragonfly+ \cite{DragonflyPlus}. Slimfly cuts a group's switch layer, effectively uniting all of its hosts with its only switch, sitting on the global layer. Slimfly's property is that any two switches on the global layer are at distance of at most two from each other. The property allows us to never need to change the switch when arriving in an intermediate group. Essentially, Slimfly has no equivalent handicap similar to Dragonfly's need to differentiate 1- and 2-indirect routes.

\begin{figure}[H]
    \centering
    \includegraphics[width=1\linewidth]{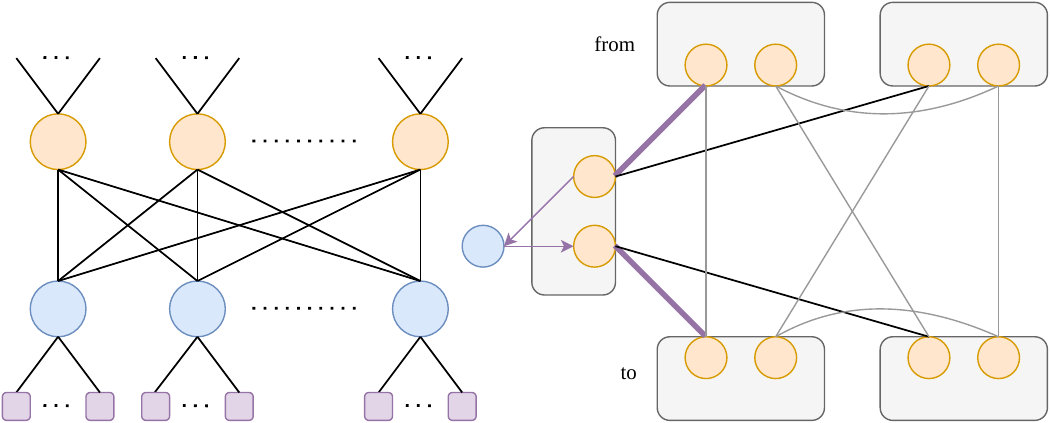}
    \caption{As an opposite, Dragonfly+ creates another layer with switches in a group. These are now divided into leaf and spine switches.}
    \label{introDfp}
\end{figure}

Leaves communicate with hosts and spines, while spines communicate with leaves and other groups' spines. Leaves and spines are united through a fully-connected layer.\\

While Dragonfly+ increases the number of hosts and the number of paths between two hosts for a given radix compared to Dragonfly, it also increases the penalty for going to the incorrect outgoing spine switch in an intermediate group from one to two hops: see the right part of figure \ref{introDfp}.\\

We are interested in the permutation oversubscription proof in expectation from the Dragonfly+ article. While it achieves the correct result, it has several flaws, and can be applied for only a subset of permutation patterns (all host pairs must be in different groups and any group must receive flows from at least two other groups). This prompts us to find a general expected value oversub ratio proof for any permutation pattern. Surprisingly, the resulting proof can be adapted to obtain approximations for any pattern. We also analyze the permutation oversub ratios without the expectation assumption. We obtain high probability theoretical bounds that are useful for small switch radixes, and compare them against results from simulations.\\

All experiments are run on the Ultra Ethernet Transport branch \cite{UEThtsim} of htsim \cite{NDP}, which aims to match as closely as possible the recent UET specification \cite{UET}.\\

We present below known oversubscription ratios on permutation patterns for popular low-diameter topologies. All rows but the last are from \cite{DragonflyPlus}'s Table I.

\begin{table}[H]
\begin{tabular}{|l|l|l|l|}
\hline
\textbf{Topology name} & \textbf{Diameter} & \textbf{\begin{tabular}[c]{@{}l@{}}Max host count\\ (radix = 36)\end{tabular}} & \textbf{\begin{tabular}[c]{@{}l@{}}Permutation\\ oversub\\ ratio\\ (expectation)\end{tabular}} \\ \hline
Dragonfly+ & 3 & 105'300 & 2:1 \\ \hline
Dragonfly & 3 & \begin{tabular}[c]{@{}l@{}}29'412\\ (wrongly reported\\ as 26'406)\end{tabular} & \begin{tabular}[c]{@{}l@{}}12:1\\ improved to\\ $\sim 2.38:1$ in \cite{DragonflyVLoversubImprovement} \end{tabular} \\ \hline
\begin{tabular}[c]{@{}l@{}}Fat Tree\\ (3-level, 2:1\\ blocking ratio)\end{tabular} & 4 & 15'552 & 2:1 \\ \hline
\begin{tabular}[c]{@{}l@{}}Fat Tree\\ (3-level, non-\\ blocking)\end{tabular} & 4 & 11'664 & 1:1 \\ \hline
Slimfly & 2 & 6'144 & 2:1 \\ \hline
HyperX (2D) \cite{HyperX} & 2 & 2'028 & 2:1 \\ \hline
\end{tabular}
\end{table}

Apart from other low diameter topologies, Dragonfly+ stands out because of the intermediate-group spine correction cost it incurs, being the most affected by it, with two hops needed to arrive to the outgoing spine, compared to one for Dragonfly and none for Slimfly.\\

Throughout the article, section \ref{section_problem_definition} provides some problem description for the 2:1 permutation oversubscription proof in expectation. Section \ref{section_dfp_proof_discussion} looks at the original proof from \cite{DragonflyPlus} and shows its shortcomings, justifying the need for a general case proof, shown in section \ref{section_oversub}. Ours can be used for other arbitrary traffic patterns to derive expected oversub ratios.\\

Section \ref{section_oversub_high_proba_bounds} computes oversubscription bounds that happen with high probability for low radixes in permutation patterns in the general case, not in expectation. This section also covers experimental results which show that the difference in FCTs created by the global topology wiring's Z matrix diminishes with an increase in radix. The Z matrix is first encountered in the oversub proof. An entry $z_{i, j}$ counts for some groups $i$, $j$ how many intermediate groups contain both $i$, $j$ in the same spine (essentially how many 1-hop corrections are available for each pair of groups).\\

We end with the conclusion in section \ref{section_conclusion}.

%% file: tex_parts/problem_definition.tex
\section{Problem definition}\label{section_problem_definition}

We follow with some introductory notation. All switches in the Dragonfly+ max-host topology have radix $k$. A group has $l$ leaf and $s$ spine switches. Each leaf has $p$ hosts. Each spine has $h$ global links attached. Generally, $k$ is even, and we use the rule-of-thumb $k/2 = h = s = l = p$. We will later see that at least for our oversubscription proof, we need to enforce this rule-of-thumb. Along the article, we also note $h = k/2$ as the half radix where the rule-of-thumb is enforced. The number of groups in the topology is $|G| = h^2 + 1$. The total number of hosts is $\totalhosts = |G| h^2 = h^2(h^2 + 1)$.\\

There are multiple types of paths available between two hosts, out of which we will use the following five (we will denote a path passing through a group-local link by $L$, and a global link by $G$):

\begin{table}[H]
\begin{tabular}{|l|l|l|}
\hline
\textbf{type} & \textbf{description} & \textbf{\# paths} \\ \hline
$LL$ ($L^2$) & group-local path that only needs to go & $1$ \\
& through a leaf switch & \\ \hline
$L^4$ & group-local path that needs to visit & $\frac{k}{2} = h$ \\
& a spine switch as well & \\ \hline
$L^2GL^2$ & global path that uses the direct global link & $1$ \\ \hline
$L^2G^2L^2$ & global path that uses an intermediate & $h - 1$ \\
& group, but only passes through a spine & (expected) \\
& (1-indirect global) & \\ \hline
$L^2GL^2GL^2$ & global path that passes through two spine & $h^2(h-1)$ \\
& switches and a leaf switch in the & (expected) \\
& intermediate group (2-indirect global) & \\ \hline
\end{tabular}
\end{table}

We will prove the number of paths for each of the five types.

For $L^2$, there is only one path, since we never leave the leaf switch that both the sender and receiver hosts share.

For $L^4$: by considering the Fully-Connected links between the leaf and spine switch layers in figure \ref{introDfp}, we notice that we can use any of the spine switches as intermediates to get from a leaf switch to another, meaning that we have $h$ distinct paths if the hosts share the group, but not the leaf switch.

For $L^2GL^2$: since there is a unique link between the two groups, we are also forced to pass through specific spine switches in both groups, meaning that there is only one distinct path.\\

We need the following property before continuing with the last two path counts.

\begin{property}
    \label{property_ab_objects_b_sectors}
    We have $a \cdot b$ objects that we want to uniformly split into $b$ sectors, each sector taking $a$ objects each. Then the probability that two specific objects end in the same sector is $(a-1) / (ab - 1)$.
\end{property}

\begin{figure}[H]
    \centering
    \includegraphics[width=0.8\linewidth]{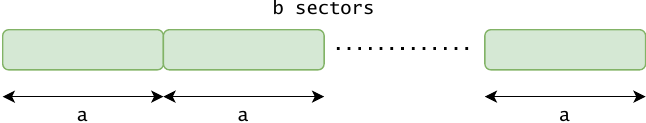}
    \caption{Fix WLOG the position of the first object. Then the second object has to randomly choose one of the $ab - 1$ left positions, out of which only $a - 1$ are in the same sector as the first object.}
    \label{problemDefinitionSectors}
\end{figure}

\begin{notation}
    Let $\group: [0, \totalhosts) \rightarrow [0, |G|)$ be a function mapping a host id to its respective group id.
\end{notation}

\begin{property}
    If we take the expectation over all possible global layer wirings, there are $h - 1$ $L^2G^2L^2$ paths on average.
\end{property}

\begin{figure}[H]
	\centering
	\includegraphics[width=0.4\linewidth]{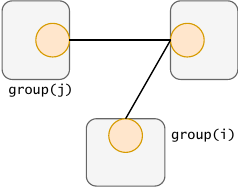}
	\caption{For a 1-indirect path between two hosts $i$ and $j$, we need to pick an intermediate group that has a spine with links to both $\group(i)$ and $\group(j)$.}
	\label{problemDefinitionFigL2G2L2}
\end{figure}

We apply {\bf property \ref{property_ab_objects_b_sectors}} on an arbitrary switch, which has $|G| - 1 = ab$ pipes leaving it, which are grouped into sectors of $h = b$ pipes each. Therefore, the probability of choosing a switch if two specific links ($\rightarrow \group(i), \rightarrow \group(j)$) are in the same spine is:

\eqalign {
    & p = \frac{\frac{|G| - 1}{h} - 1}{|G| - 2} = \frac{\frac{h^2}{h} - 1}{|G| - 2}
}

And the expected number of 1-indirect groups is $p \cdot (|G| - 2) = h - 1$.\\

\begin{property}
    The expected number of $L^2GL^2GL^2$ paths is $h^2(h-1)$.
\end{property}

The expected number of groups where we can't directly jump between $\group(i)$ and $\group(j)$ is:

\eqalign {
    & (1-p) \cdot (|G|-2) = |G|-2 - \frac{|G|-1}{h} + 1 = h(h - 1)
}

In the interior of any intermediate group we can choose any of the $h$ leaf switches between the two fixed spines, so the expected amount of distinct paths is $h^2(h-1)$.\\

%% file: tex_parts/dfp_proof_discussion.tex
\section{Original oversubscription proof discussion}\label{section_dfp_proof_discussion}

We analyze \cite{DragonflyPlus}'s own average-case 2:1 oversubscription proof for the permutation pattern \footnote{It can be found in the article in section \RNum{5}, subsection D.}. We will show the two mistakes that cancel out to produce the correct ratio and discuss the observation with no proof that was implied.

\begin{assumption}
    Throughout the paper, we will assume a simple oblivious packet sprayer i.e. Valiant \cite{Valiant}. It picks for each packet a random intermediate group first, that generally describes a non-min-cost path. Afterwards, it follows any min-cost route from it, with tie-breaks chosen randomly.
\end{assumption}

\begin{assumption}
    \label{assumption_og_send_recv_pairs}
    The original proof only considers worst-case permutations, in which all sender-receiver host pairs must be in different groups.
\end{assumption}

\begin{figure}[H]
    \centering
    \includegraphics[width=0.5\linewidth]{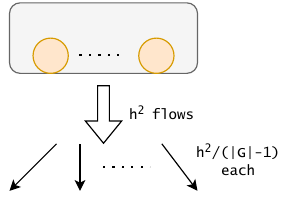}
    \caption{As a result, $h^2$ flows must leave a group.}
    \label{problemDefinitionDfpProof1}
\end{figure}

Since we can use intermediate groups, any packet is allowed to choose any outgoing pipe, resulting in a $h^2 / (|G| - 1) = 1$ expected load per global edge leaving the group.

\begin{figure}[H]
    \centering
    \includegraphics[width=\linewidth]{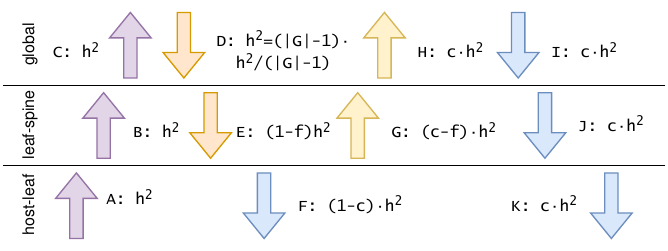}
    \caption{A diagram summarizing the coexisting flows in a group specific to the proof. We color traffic that is still in the sender group as purple, receiver group as blue, intermediate group as yellow, and unknown (receiver or intermediate) as orange.}
    \label{problemDefinitionDfpProof2}
\end{figure}

Since all global level links have $h^2 / (|G| - 1)$ expected load (before accounting for traffic leaving intermediate groups) and we have $|G| - 1$ global links entering in a group, the expected incoming total traffic is $h^2$ for step D.\\

\begin{notation}
    Some flow from $D$ doesn't need to travel to the leaf layer, i.e. the 1-indirect flow. Let its expected fraction be $f \cdot h^2$.    
\end{notation}

\begin{notation}
    Out of the $(1-f)h^2$ direct and 2-indirect flow reaching the leaves, we expect only $(1 - c)h^2 = 1 / (|G| - 1) \cdot h^2 = 1$ unit of traffic (where $c = 1 - 1/h^2$) to have been correctly sent to the destination group (step F).
\end{notation}

As a result, the other $(c-f) \cdot h^2$ is 2-indirect (step G). From now on, it along with the $f \cdot h^2$ 1-indirect flow must follow min-cost \emph{paths} to their destinations.

\begin{observation}
    \label{observation_follow_mincost}
    The distinction between a flow following min-cost \emph{paths} and \emph{path} is very important. A packet by itself must now follow a min-cost path, but a flow is made out of multiple packets, and so it follows min-cost paths instead.
\end{observation}

\begin{mistake}
    The original proof fails to account the $(1 - c) \cdot h^2 > 0$ flow that is direct and that is not sent back up into the global layer. If this were the only oversight of the original proof, the oversubscription factor would be strictly less than $2:1$, although it would converge to $2:1$ as $h \rightarrow +\infty$.
\end{mistake}

\begin{mistake}
    The second, bigger mistake with the proof occurs at step H. To get an expected load per outgoing global link, we should divide $c \cdot h^2$ by how many links we can actually use out of $h^2$, but it's not trivial to compute this, and the proof wrongly assumes that the answer is $h^2$ itself. For each packet, we certainly can't use the link it arrived through in the intermediate group, since it would send it back to the receiver, so we can divide by at most $h^2-1$.
\end{mistake}

\begin{property}
    \label{property_two_groups_send_one}
    In order to divide by $h^2 - 1$, we need for each receiver group to have two hosts in distinct groups sending to it.
\end{property}

\begin{figure}[H]
    \centering
    \includegraphics[width=0.4\linewidth]{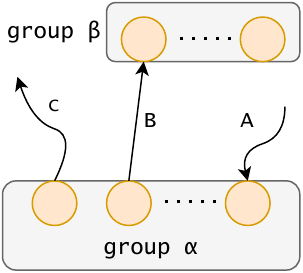}
    \caption{Suppose by contradiction that there is exactly one group $\alpha$ in the permutation sending traffic to a group $\beta$.}
    \label{problemDefinitionDfpProof3}
\end{figure}

\begin{observation}
    Because of {\bf assumption \ref{assumption_og_send_recv_pairs}}, there must be at least one group that sends to $\beta$.
\end{observation}

Then the pipe marked with B in figure \ref{problemDefinitionDfpProof3} cannot carry indirect traffic, since we would need $\alpha$ to be an indirect group for a pair $(\gamma \neq \alpha, \beta)$, but the group $\gamma$ cannot exist because of the supposition. Any packet for which $\alpha$ acts as indirect, e.g. one that arrived through pipe A cannot leave through it, nor through pipe B, so we can divide at most by $h^2 - 2$ for group $\alpha$ (e.g. pipe C can be counted).\\

On the contrary, if the contradiction is not fulfilled and we allow $\gamma$ to exist, then if we send flow for enough time, pipe B can eventually be used. All other pipes $\delta \rightarrow \beta$ with $\delta \neq \alpha$ already have indirect packets on them since $\delta$ acts as an intermediate group for $(\alpha, \beta)$. This finishes the proof for {\bf property \ref{property_two_groups_send_one}}, and should offer some insight on {\bf observation \ref{observation_follow_mincost}}.\\

Finishing figure \ref{problemDefinitionDfpProof2}, in step I we will get the $c \cdot h^2$ which are for the destination group, which then gets propagated through steps J and K. As a result, the expected leaf-spine/spine-spine up/down link oversubscription is $h^2/h^2 + ch^2/(h^2-1): 1 = 2:1$.

%% file: tex_parts/2_1_oversub.tex
\section{Oversubscription general proof}\label{section_oversub}

We have seen in section \ref{section_dfp_proof_discussion} that the original proof only stands for permutation patterns, where all host pairs are from different groups and there are at least two sender groups for each receiver group. This is a strong enough incentive to look for a general-case proof that fixes the original proof's shortcomings.\\

While the original proof holds only for a subset of the permutation pattern, we will support a superset of them: we only constrain all hosts to send and receive at most line rate. This can be further relaxed to obtain approximate expected FCT estimates for any traffic pattern.\\

Supporting any pattern in which all hosts must receive and send at most line rate can be expressed in the following property:

\begin{property}
    \label{property_rate_sum}
    Let any link from the topology be composed of two opposed one-way sub-links. Let $\amt_{i, j} \in [0, 1]$ be the fraction of the one-way sub-link bandwidth that host $i$ sends to host $j$ in a timestep.
\end{property}

Then:

\eqalign {
    & \sum_{j=1}^{\totalhosts} {\amt}_{i, j} \leq 1 \text{ (send at most line rate to host } i \text{'s leaf switch)}\\
    & \sum_{j=1}^{\totalhosts} {\amt}_{j, i} \leq 1 \text{ (receive at most line rate from host } i \text{'s leaf)}
}

These can be written literally as $\sum_{i \text{ or } j} W \amt_{i, j} \leq W$, where $W$ is the amount of packets a host can send without feedback, i.e. the congestion window (CWND) variable. We consider $\amt_{i, i} = 0$.

\begin{section_overview}
    We want to prove that regardless of the values that the matrix $\amt$ contains, as long as it fulfills {\bf property \ref{property_rate_sum}}, then the expected flow sums for all other network sub-links don't exceed $2$ (i.e. in order to not pass expected 100\% bandwidth usage on leaf-spine and global links, we need to not pass 50\% usage on host-leaf links).
\end{section_overview}

\begin{observation}
    Esentially, the $\amt$ matrix is an input to a Linear Program. We prove that the worst $\amt^*$ matrix that an optimizer would find will give each flow value passing through a leaf-spine, or a global link a maximum expected value of at most $2$. The optimizer works with the coefficients that split the traffic given by Valiant routing. The optimizer maximizes the maximum \footnote{In order to not resort to a Mixed-Integer Linear Program to maximize the maximum flow, we can instead call $\Theta(h^4)$ LPs, each maximizing one leaf-spine or global link flow. Because of the implication of {\bf property \ref{property_uniform_expected_load_increase}}, the Dragonfly+ topology is symmetrical in expectation, and so we only need to call $2$ LPs instead of $\Theta(h^4)$, one for an arbitrary leaf-spine link and one for a global link.} flow value passing through a leaf-spine, or a global link.
\end{observation}

\begin{question}
    Why not just solve the two LPs for practical values of $h$ instead of obtaining a link flow upper bound? The $\amt$ matrix has $\Theta(h^8)$ entries. Assuming that $\amt^*$ is dense and that matrix multiplication can be done in quadratic time, a naive application of an efficient LP solver could be $O(h^{16 + \epsilon})$.
\end{question}

\begin{observation}
    The permutation pattern satisfies {\bf property \ref{property_rate_sum}}, as for any host $i$ is an unique host $\text{pe}(i)$ that concentrates host $i$'s entire line rate: $\amt_{i, \text{pe}(i)} = 1$.
\end{observation}

\input{tex_parts/2_1_oversub_part_1}

\input{tex_parts/2_1_oversub_part_2}

\input{tex_parts/2_1_oversub_discussion}

%% file: tex_parts/2_1_oversub_part_1.tex
\subsection{Global Sub-Link Flow Expected Bound}\label{subsection_oversub_global}

\begin{property}
    Let $1 \leq i, j \leq \totalhosts$ be two hosts from different groups. Then the flow that $i \rightarrow j$ imposes on the global layer is $W \cdot \amt_{i, j}$.
\end{property}

\begin{notation}
    \label{notation_z}
    Let $z_{\group(i), \group(j)}$ be the number of groups whose connections with groups $\group(i)$ and $\group(j)$ are in the same spine. We may note $(z \circ \group)_{i, j} = z_{\group(i), \group(j)}$.
\end{notation}

\begin{notation}
    Let $(\#0)_{i, j}$, $(\#1)_{i, j}$, $(\#2)_{i, j}$ be random variables counting how many packets on the $i \rightarrow j$ host-to-host flow get sent on the only direct link, one of the $z_{\group(i), \group(j)}$ 1-indirect links, or one of the $|G| - 2 - z_{\group(i), \group(j)}$ 2-indirect links:
    
    \eqalign {
        & (\#0)_{i, j} + (\#1)_{i, j} + (\#2)_{i, j} = W \cdot {\amt}_{i, j}\\
        & (\#0)_{i, j} \sim \text{ Binomial} \bigparen{W \cdot {\amt}_{i, j},\, \frac{1}{h^2}}\\
        & (\#1)_{i, j} \sim \text{ Binomial} \bigparen{W \cdot {\amt}_{i, j},\, \frac{z_{\group(i), \group(j)}}{h^2}}\\
        & (\#2)_{i, j} \sim \text{ Binomial} \bigparen{W \cdot {\amt}_{i, j},\, \frac{h^2 - 1 - z_{\group(i), \group(j)}}{h^2}}
    }
\end{notation}

i.e. $(\#0)_{i, j}$ is the sum of $W \cdot {\amt}_{i, j}$ simple Bernoulli events, each with success probability $1 / h^2$, since only one out-queue is a direct link to $\group(j)$.\\

\begin{notation}
    Let $\alpha_{i, j}$ represent the $i \rightarrow j$ flow fraction going directly to $j$'s group. Similarly, let $\beta_{i, j}$ and $\gamma_{i, j}$ represent the flow fractions going on 1-indirect and 2-indirect routes to $j$'s group.
\end{notation}

\eqalign {
    & \mathbb{E}(\alpha_{i, j}) = \mathbb{E} \bigparen{ \frac{(\#0)_{i, j}}{W \cdot {\amt}_{i, j}} } = \frac{\mathbb{E}((\#0)_{i, j})}{W \cdot {\amt}_{i, j}} = \frac{1}{h^2} = \frac{4}{k^2} =^\text{not} \alpha\\
    & \mathbb{E}(\beta_{i, j}) = \frac{(z \circ \group)_{i, j}}{h^2} = \alpha \cdot (z \circ \group)_{i, j}\\
    & \phantom{\mathbb{E}(\beta_{i, j}) = \,} \text{ (so expected } \alpha \text{ per each of } (z \circ \group)_{i, j} \text{ edges)}\\
    & \mathbb{E}(\gamma_{i, j}) = \frac{h^2 - 1 - (z \circ \group)_{i, j}}{h^2} = \alpha \cdot (h^2 - 1 - (z \circ \group)_{i, j})\\
    & \phantom{\mathbb{E}(\gamma_{i, j}) = \,} \text{ (exp. } \alpha \text{ per edge)}\\
}

\begin{property}
    \label{property_uniform_expected_load_increase}
    Valiant routing increases load uniformly (in \emph{expectation}) over all affected global edges for any global flow $i \rightarrow j$. Each load increases with $\alpha \cdot \amt_{i, j}$.
\end{property}

\begin{figure}[H]
	\centering
	\includegraphics[width=0.7\linewidth]{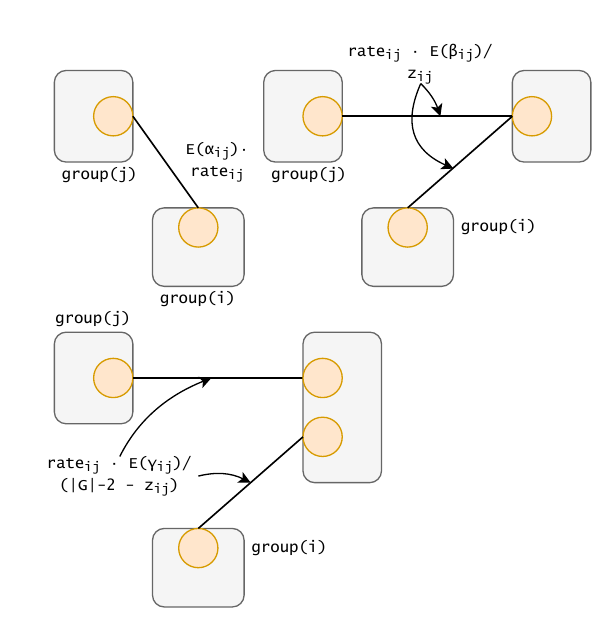}
	\caption{{\bf Property \ref{property_uniform_expected_load_increase}} drawn. $z_{\group(i), \group(j)}$ vanishes when we consider the expected load.}
	\label{figEqualGlobalLoad}
\end{figure}

We will now compute the \emph{expected} load on the two global sub-links that unite group $x$ to group $y$. Let $g_{a, b}$ represent the id of the $b$-th host from the $a$-th group ($1 \leq a \leq |G|, 1 \leq b \leq lp$).

\begin{enumerate}[i]
    \item The load from $L^2GL^2$ paths:

    \begin{figure}[H]
    	\centering
    	\includegraphics[width=0.4\linewidth]{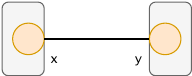}
    	\caption{Every pair of hosts' flows from groups $x$ and $y$ pass through this edge.}
    	\label{figGlobalBoundL2GL2}
    \end{figure}

    The load for packets from $x$ to $y$, only adding to the $x \rightarrow y$ sub-link:

    \eqalign {
        & \sum_{i=1}^{lp}\sum_{j=1}^{lp} \alpha \cdot {\amt}_{g_{x, i}, g_{y, j}}
    }
    The load for the $y \rightarrow x$ sub-link:

    \eqalign {
        & \sum_{i=1}^{lp}\sum_{j=1}^{lp} \alpha \cdot {\amt}_{g_{y, j}, g_{x, i}}
    }
    
    From now on, we will forgo mentioning the load for the $y \rightarrow x$ sub-link, since if $\amt_{i, j}$ appears in the $x \rightarrow y$ load sum, $\amt_{j, i}$ will appear in the $y \rightarrow x$ load sum, so both sub-links will be eventually bounded to the same limit.\\

    \item The load from $L^2G^2L^2$ paths:

    \begin{figure}[H]
    	\centering
    	\includegraphics[width=0.7\linewidth]{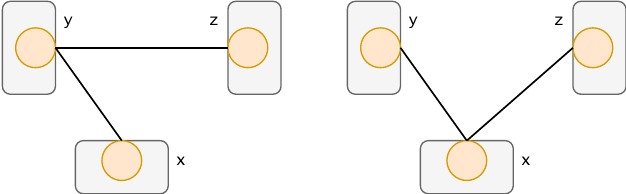}
    	\caption{Both $x$ and $y$ can act as intermediate groups.}
    	\label{figGlobalBoundL2G2L2}
    \end{figure}

    For $y$ as an intermediate group:
    
    \eqalign {
        & \sum_{\substack{z \in A\\|A| = h-1}} \sum_{i=1}^{lp} \sum_{t=1}^{lp} \alpha \cdot {\amt}_{g_{x, i}, g_{z, t}}
    }
    
    For $x$ as an intermediate group:

    \eqalign {
        & \sum_{\substack{z \in B\\|B| = h-1}} \sum_{t=1}^{lp} \sum_{j=1}^{lp} \alpha \cdot {\amt}_{g_{z, t}, g_{y, j}}
    }

    \item The load from $L^2GL^2GL^2$ paths:

    \begin{figure}[H]
    	\centering
    	\includegraphics[width=0.7\linewidth]{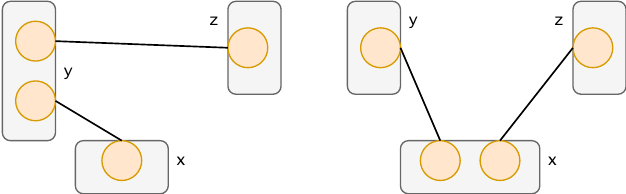}
    	\caption{Again, both $x$ and $y$ can be intermediates.}
    	\label{figGlobalBoundL2GL2GL2}
    \end{figure}

    For $y$ as an intermediate group:

    \eqalign {
        & \sum_{\substack{z \in C\\|C| = h(s-1)}} \sum_{i=1}^{lp} \sum_{t=1}^{lp} \alpha \cdot {\amt}_{g_{x, i}, g_{z, t}}
    }

    For $x$ as an intermediate group:

    \eqalign {
        & \sum_{\substack{z \in D\\|D| = h(s-1)}} \sum_{t=1}^{lp} \sum_{j=1}^{lp} \alpha \cdot {\amt}_{g_{z, t}, g_{y, j}}
    }
\end{enumerate}

\begin{property}
    \label{property_disjoint_sets}
    Since there is exactly one link uniting any two different groups, no group can be found twice in $A$, $B$, $C$, or $D$. Also, $A \cap C = \emptyset$, since by changing the outgoing spine from $y$ we cannot reach any groups that were reachable through the ingoing spine in $y$. Similarly, $B \cap D = \emptyset$. Also, from {\romannumeral 2} and {\romannumeral 3}, $x, y \notin A \cup B \cup C \cup D$, since in {\romannumeral 2} we don't take the link that we just came through, and in {\romannumeral 3} we change the outgoing spine.
\end{property}

Using $A \cap C = \emptyset$ from {\bf property \ref{property_disjoint_sets}}, we will unite the loads from {\romannumeral 1}, {\romannumeral 2} (first part), and {\romannumeral 3} (first part). Afterwards, we will apply {\bf property \ref{property_rate_sum}} with the fixed hosts $g_{x, i}$:

\eqalign {
    & \alpha \sum_{i=1}^{lp} \bigparen{\sum_{j=1}^{lp} {\amt}_{\underline{g_{x, i}}, g_{y, j}} + \sum_{z \in A \cup C} \sum_{t=1}^{lp} {\amt}_{\underline{g_{x, i}}, g_{z, t}}}
}

From {\bf property \ref{property_disjoint_sets}}, $y \notin A \cup C \Rightarrow g_{y, j} \neq g_{z \in A \cup C, t}$. Also, $A \cap C = \emptyset \Rightarrow g_{z \in A, t} \neq g_{z \in C, t}$. This means that our previous sum is at most:

\eqalign {
    & \alpha \sum_{i=1}^{lp} \sum_{j=1}^{\totalhosts} {\amt}_{\underline{g_{x, i}}, j} \applypropone \alpha \sum_{i=1}^{lp} 1 = \alpha \cdot lp = \frac{4}{k^2} \cdot \frac{k}{2}\frac{k}{2} = 1
}

Similarly, we will apply {\bf property \ref{property_rate_sum}} (the second inequality this time) over the fixed hosts $g_{y, j}$ using $B \cap D = \emptyset$ from {\bf property \ref{property_disjoint_sets}} with the remaining loads from {\romannumeral 2} and {\romannumeral 3}, obtaining another flow bound of one.\\

Summing both flow bounds gives us an expected total bound for the $x \rightarrow y$ sub-link of at most $1 + 1 = 2$. Implicitly, we can get a bound of $2$ as well for the $y \rightarrow x$ sub-link.

%% file: tex_parts/2_1_oversub_part_2.tex
\subsection{Leaf - Spine Sub-Link Flow Expected Bound}\label{subsection_oversub_local}

We will apply the same rationale as in the previous subsection, aiming to apply {\bf property \ref{property_rate_sum}} as little times as possible, although bounding may be more complicated. For a fixed arbitrary group, let $\leaf_{x, q}$ represent the host id of the $q$-th host under the $x$-th leaf router ($1 \leq x \leq l, 1 \leq q \leq p$).\\

We will note the leaf router with $x$ and the spine router with $y$.\\

\begin{enumerate}[i]
    \setcounter{enumi}{3}

    \item We first count the load from $L^4$ paths.
    
    \begin{figure}[H]
        \centering
        \includegraphics[width=0.5\linewidth]{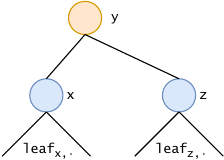}
        \caption{The receiver host will be in the same group, under leaf $z$.}
        \label{figLocalBoundL4}
    \end{figure}

    The path is obligated to pass through $x$ since it has one end in one host under it, but it doesn't necessarily need to pass through $y$. It can choose any spine router and continue to the $t$ leaf router. Because of oblivious spraying, on average only $1/s$ will pass through $x \rightarrow y$:

    \eqalign {
        & \frac{1}{s} \sum_{i=1}^p \sum_{\substack{t=1\\t \neq x}}^l \sum_{j=1}^p {\amt}_{\leaf_{x, i}, \leaf_{t, j}}
    }
\end{enumerate}

We will count in {\romannumeral 5} the load imposed by path types $L^2GL^2, L^2G^2L^2, L^2GL^2GL^2$ (we will first only consider $L^2GL^2GL^2$ paths where the $x \rightarrow y$ sub-link is not in the intermediate group).

\begin{figure}[H]
    \centering
    \includegraphics[width=0.5\linewidth]{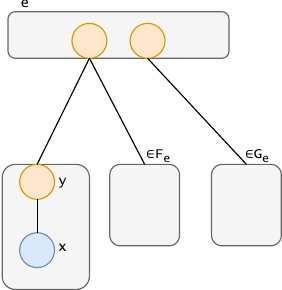}
    \caption{We will leave $x \rightarrow y$'s group through one of the $h$ links outgoing from $y$. We will arrive in a group $e \in E, |E| = h$.}
    \label{figBoundAllNonIntermediate}
\end{figure}

For a fixed $e \in E$, this group can either be final (accounting for $L^2GL^2$ paths) or intermediate, and we can leave it through the same spine that we arrived in (arriving in a group from $F_e, |F_e| = h-1$, accounting for $L^2G^2L^2$ paths), or through another spine (arriving in a group from $G_e, |G_e| = (s-1)h$, accounting for some $L^2GL^2GL^2$ paths).

\begin{property}
    \label{property_disjoint_sets_local}
    For a fixed $e \in E$, no two groups can be found twice in $F_e$ or $G_e$, and $F_e \cap G_e = \emptyset$ since there is only one link between any two different groups, and we get to groups in $F_e$ and $G_e$ by leaving through different spines. Also, $e \notin F_e \cup G_e$, since we must leave $e$ to get to $F_e \cup G_e$. For $H_e = \{e\} \cup F_e \cup G_e$, it follows that $|H_e| = 1 + h-1 + (s-1)h = sh$.
\end{property}

\begin{enumerate}[i]
    \setcounter{enumi}{4}

    \item The load is:

    \eqalign {
        & \alpha \sum_{e \in E} \sum_{z \in H_e} \sum_{i=1}^p \sum_{j=1}^{lp} {\amt}_{\leaf_{x, i}, g_{z, j}}
    }
\end{enumerate}

We will now unite the loads from {\romannumeral 4} and {\romannumeral 5}. We will use $\alpha = 1/s^2$ and factor $1/s^2$ out of everything:

\eqalign {
    & \frac{1}{s^2} \sum_{i=1}^p \bigparen{s \sum_{\substack{t=1\\t \neq x}}^l \sum_{j=1}^p {\amt}_{\leaf_{x, i}, \leaf_{t, j}} +\\
    & \phantom{\frac{1}{s^2} \sum_{i=1}^p \,\,\,} \sum_{e \in E} \sum_{z \in H_e} \sum_{j=1}^{lp} {\amt}_{\leaf_{x, i}, g_{z, j}}}
}

Since the first double sum in the parenthesis is repeated $s$ times, and $s = h = |E|$, we will distribute one sum for each $e \in E$:

\eqalign {
    & \frac{1}{s^2} \sum_{i=1}^p \sum_{e \in E} \bigparen{
    \sum_{\substack{t=1\\t \neq x}}^l \sum_{j=1}^p {\amt}_{\leaf_{x, i}, \leaf_{t, j}} + \sum_{z \in H_e} \sum_{j=1}^{lp} {\amt}_{\leaf_{x, i}, g_{z, j}}}
}

We can swap the first two summation signs ($i, e$) and apply {\bf property \ref{property_rate_sum}} over the fixed hosts $\leaf_{x, i}$ in the parenthesis, since $\leaf_{t, j}$ is part of $x \rightarrow y$'s group which cannot be found in $H_e$. We also use {\bf property \ref{property_disjoint_sets_local}} to guarantee that no $g_{z, j}$ is found twice:

\eqalign {
    & \frac{1}{s^2} \sum_{e \in E} \sum_{i=1}^p \bigparen{
    \sum_{\substack{t=1\\t \neq x}}^l \sum_{j=1}^p {\amt}_{\underline{\leaf_{x, i}}, \leaf_{t, j}} +\\
    & \phantom{\frac{1}{s^2} \sum_{e \in E} \sum_{i=1}^p \,} \sum_{z \in H_e} \sum_{j=1}^{lp} {\amt}_{\underline{\leaf_{x, i}}, g_{z, j}}} \leq\\
    & \leq \frac{1}{s^2} \sum_{e \in E} \sum_{i=1}^p \sum_{j=1}^{\totalhosts} {\amt}_{\underline{\leaf_{x, i}}, j} \applypropone
    \frac{1}{s^2} \sum_{e \in E} \sum_{i=1}^p 1 = \frac{ph}{s^2} = 1
}

\begin{enumerate}[i]
    \setcounter{enumi}{5}
    
    \item We finish by counting the load imposed by $L^2GL^2GL^2$ paths for which $y \rightarrow x$ is part of the intermediate group.

    \begin{figure}[H]
        \centering
        \includegraphics[width=0.6\linewidth]{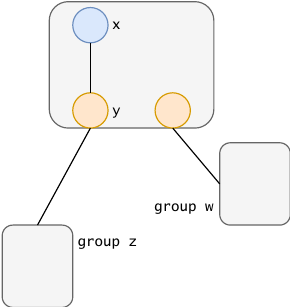}
        \caption{The load going from group $z$ to group $w$ enters $y \rightarrow x$'s group. The ingoing spine is fixed in $y$. $y \rightarrow x$ will support on average $1/l$ of the total load, since we can spray over any leaf router in the intermediate group.}
        \label{figBoundIntermediate}
    \end{figure}

    The load on the $y \rightarrow x$ sub-link is:

    \eqalign {
        & \frac{\alpha}{l} \sum_{\substack{z \in I\\|I| = h}} \,\,\, \sum_{\substack{w \in J\\|J| = (s-1)h}} \sum_{i=1}^{lp} \sum_{j=1}^{lp} {\amt}_{g_{z, i}, g_{w, j}}
    }
\end{enumerate}

\begin{property}
    \label{property_disjoint_sets_local2}
    No two groups can be found twice in either $I$ or $J$, and $I \cap J = \emptyset$, similarly like {\bf property \ref{property_disjoint_sets_local}}.
\end{property}

We apply {\bf property \ref{property_rate_sum}} over the fixed hosts $g_{z, i}$:

\eqalign {
    & \phantom{\leq \,\,\,\,} \frac{\alpha}{l} \sum_{i=1}^{lp} \sum_{z \in I} \bigparen{\sum_{w \in J} \sum_{j=1}^{lp} {\amt}_{\underline{g_{z, i}}, g_{w, j}}} \leq\\
    & \leq \frac{\alpha}{l} \sum_{i=1}^{lp} \sum_{z \in I} \sum_{j=1}^{\totalhosts} {\amt}_{\underline{g_{z, i}}, j} \applypropone \frac{\alpha}{l} \sum_{i=1}^{lp} \sum_{z \in I} 1 = \frac{\alpha \cdot lph}{l} = 1
}

Implicitly, we get a bound of $1$ from {\romannumeral 6} for the $x \rightarrow y$ sub-link as well. We sum the flow bounds from {\romannumeral 4}, {\romannumeral 5} and {\romannumeral 6}, getting an expected upper bound for any leaf-spine sub-link of at most $1 + 1 = 2$.

%% file: tex_parts/2_1_oversub_discussion.tex
\subsection{Discussion}\label{subsection_oversub_discussion}

\begin{theorem}
    We have shown that it is possible for any $\amt$ matrix that supports {\bf property \ref{property_rate_sum}} to bound all (global and leaf-spine) sub-link flows in expectation to at most two times the maximum flow on the host-leaf sub-links, making any leaf and spine router's oversubscription ratio at most 2:1 against the host-leaf flow in expectation.
\end{theorem}

\begin{observation}
    The multiple constraints that we have encountered during subsections \ref{subsection_oversub_global} ($\alpha lp = 1$) and \ref{subsection_oversub_local} ($\alpha s^2 = 1$, $s = h$, $\alpha ph = 1$) show that the general rule of thumb $h = s = l = p$ must hold in order for the presented proof to work.
\end{observation}

\begin{property}
    Instead of capping the send and receive per host up to at most line rate, if any host sends at most $s$ times the line rate, and receives at most $r$ times the line rate, then the estimate in expectation is $2\max(s, r):1$.
\end{property}

Some example estimates can be found below. The approximations perform better if $|s-r| \rightarrow 0$:

\begin{table}[H]
\begin{tabular}{|l|l|l|l|}
\hline
\textbf{Traffic Pattern} & \textbf{s} & \textbf{r} & \textbf{\begin{tabular}[c]{@{}l@{}}Oversubscription\\ ratio estimate\end{tabular}} \\ \hline
Permutation & 1 & 1 & 2:1 \\ \hline
Incast & 1 & $\totalhosts - 1$ & \begin{tabular}[c]{@{}l@{}}$2(\totalhosts - 1) : 1$\\ (instead of the correct\\ $(\totalhosts - 1) : 1$)\end{tabular} \\ \hline
All-to-All & $\totalhosts - 1$ & $\totalhosts - 1$ & $2(\totalhosts - 1) : 1$ \\ \hline
\begin{tabular}[c]{@{}l@{}} Partial All-to-All\\ $m$ groups out of $|G|$ \end{tabular} & $mh^2 - 1$ & $mh^2 - 1$ & \begin{tabular}[c]{@{}l@{}}$2(mh^2 - 1) : 1$\\ (see figure \ref{figPartialAllToAllPracticeOversub})\end{tabular}\\ \hline
\end{tabular}
\end{table}

The first of several experiments from htsim can be found below. Some relevant htsim arguments are mentioned:

\begin{table}[H]
\begin{tabular}{|l|l|}
    \hline
    \textbf{Feature Name} & \textbf{Description} \\ \hline
    Packet Size & $4$ KB + $54$ bytes \\ \hline
    Flow Size & $2 \cdot 10^6$ bytes \\ \hline
    Link Speed & $100$ Gbps \\ \hline
    Link Delay & $1 \mu$s \\ \hline
    Switch Delay & $0$ s \\ \hline
    Max RTT & $15 \mu$s \\ \hline
    Max Hops & 8 \\ \hline
    \begin{tabular}[c]{@{}l@{}}Path Count used\\ in ECMP dummy \end{tabular} & 128 \\ \hline
    Congestion Control & \begin{tabular}[c]{@{}l@{}}Sender-only\\ Receiver-only\\ no CC with infinite queue size, CWND\end{tabular} \\ \hline
    BDP & $33$ packets \\ \hline
    Queue Size & BDP \\ \hline
    ECN tagging & \begin{tabular}[c]{@{}l@{}}only if CC is used:\\ from $20\%$ to $80\%$ queue size \end{tabular} \\ \hline
\end{tabular}
\end{table}

\begin{figure}[H]
    \centering
    \includegraphics[width=0.8\linewidth]{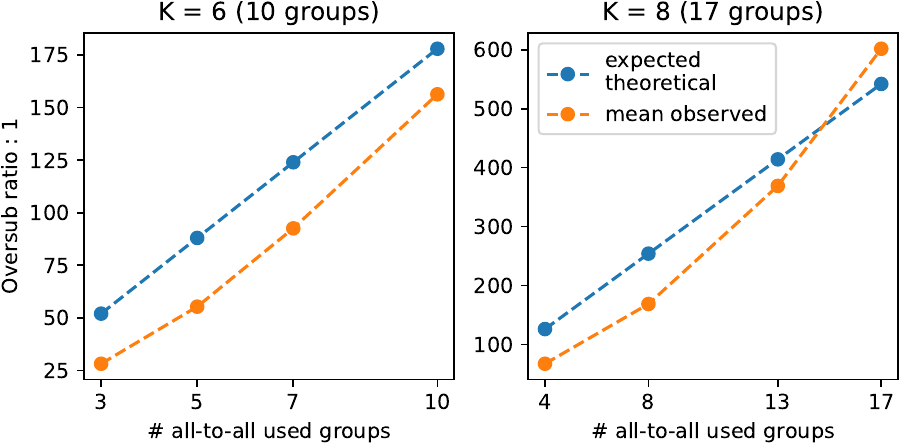}
    \caption{$m$-group partial all-to-all, sender-only CC: we compare the expected theoretical $2(mh^2 - 1) : 1$ ratio with the mean ratio observed in practice.}
    \label{figPartialAllToAllPracticeOversub}
\end{figure}

We take the mean over a batch of $m$-groups such that each group is selected in as many batches as another (Balanced Incomplete Block Design). We obtain the ratio in practice by dividing the mean FCT with the mean time needed to finish one direct global flow ($\sim 180 \mu$s). While the behaviour observed in practice isn't linear, it is approximated decently by our formula.

%% file: tex_parts/2_1_oversub_high_proba_bound.tex
\section{Oversubscription: High Probability Bounds}\label{section_oversub_high_proba_bounds}

We have proven in section \ref{section_oversub} that even a simple Valiant load balancer can achieve an expected 2:1 oversubscription bound on all switches, given that all hosts send and receive at most line rate per timestep.\\

In this section, we concentrate on finding oversubscription bounds that hold with high probability (w.h.p.) in the general case, with no expectation presumption.

\begin{section_overview}
    If we want to use the expected value proof's framework, we should simply bound $\alpha$ w.h.p. However, this is difficult to do while keeping the same general per-host line rate constraint. Remember that:
    
    \eqalign {
        & (\#0)_{i, j} \sim \text{ Binomial} \bigparen{W \cdot {\amt}_{i, j},\, \frac{1}{h^2}}
    }
    
    An immediate idea would be to apply Chernoff's inequality to upper bound $(\#0)_{i, j}$. However, the bound is very inefficient if there are very few Bernoulli events summed. Here there are $W \cdot {\amt}_{i, j}$, and we don't have any particular lower bound for ${\amt}_{i, j}$ but $0$. We concentrate on the permutation pattern, as there is an unique host $\text{pe}(i)$ that concentrates host $i$'s entire line rate: $\amt_{i, \text{pe}(i)} = 1$.
\end{section_overview}

Applying Chernoff's inequality on $(\#0)_{i, \text{pe}(i)} \sim \text{Bin}(W,\, 1/h^2)$ gives:

\eqalign {
    & P \bigparen{(\#0)_{i, \text{pe}(i)} \geq \underset{=^\text{not} \mu}{\frac{W}{h^2}}(1 + \delta)} \leq \exp \bigparen{- \frac{\delta^2}{3} \mu} \,\, \forall \, \delta \in (0, 1)
}

For each host $i$ we have $h^2$ events, one for each outgoing link out of $i$'s group $\group(i)$, so in total for $\totalhosts$ hosts we have $\totalhosts h^2$ events. By applying the union bound we obtain:

\eqalign {
    & P \bigparen{ \bigcup_{i} \bigcup_{\substack{\text{global link}\\ \in \group(i)}} (\#0)_{i, \text{pe}(i)} \geq \mu(1 + \delta)} \leq \totalhosts h^2 \exp \bigparen{- \frac{\delta^2}{3} \mu}
}

Then the probability that none of these unions happen is the complement:

\eqalign {
    & p = 1 - h^4(h^2 + 1) \exp \bigparen{- \frac{\delta^2}{3} \cdot \frac{W}{h^2}}
}

\begin{property}
    \label{property_chernoff_union_bound}
    If we use the lowest $\delta$ such that $p$ would be considered high probability, then $\alpha_{i, j} \leq (1 + \delta) / h^2$ w.h.p, and the oversubscription bound would be $2(1 + \delta): 1$ w.h.p.
\end{property}

We will compute the lowest $\delta$ s.t. $p \geq 1 - 10^{-2}$ numerically for each $h$, with $\mu = W / h^2$ and $\totalhosts = h^2(h^2+1)$. Since $W$ is present here, we need an accurate estimate of CWND.\\

If we know the congestion control mechanism, then we should follow the CWND imposed by it, usually $1$ or $1.5$ times the BDP. The largest value of CWND happens if we use infinite queue sizes and no congestion control.

\begin{property}
    Following the 2:1 expected oversubscription rate, we estimate that the largest number of concurring packets on a pipe is:
    
    \eqalign {
        & 2 \cdot \frac{\text{flow size}}{\text{packet size}} = 2 \cdot \frac{2 \cdot 10^6}{4096 + 54} \simeq 2 \cdot 481 = 962 \text{ packets}
    }    
\end{property}

For $W = 962$ packets, we get the following numerical results:

\begin{table}[H]
\begin{tabular}{|l|l|l|l|l|}
\hline
$h$ & $\totalhosts$ & $p$ & $\delta$ & $2(1 + \delta): 1$ \\ \hline
$2$ & $20$ & $0.990$ & $0.335$ & $2.67 : 1$ \\ \hline
$3$ & $90$ & $0.990$ & $0.565$ & $3.13 : 1$  \\ \hline
$4$ & $272$ & $0.990$ & $0.805$ & $3.61 : 1$ \\ \hline
$5$ & $650$ & $0.956$ & $1$ & bound is unusable for $h \geq 5$. \\ \hline
\end{tabular}
\end{table}

Unfortunately, the Chernoff bound will always be weaker with a higher $h$, as $(\#0)_{i, \text{pe}(i)} \sim \text{ Bin}(W,\, 1/h^2)$'s mean decreases when $h$ increases, equivalent with lowering the number of trials $W$ instead.\\

\begin{observation}
    We could try to be less strict and allow some event complements to not hold, instead of wanting all of them to be true. For example, we could allow $h^2$ out of $\totalhosts$ hosts to not meet the bound.
    
    \eqalign {
        & p = 1 - {\totalhosts \choose h^2} h^2 \cdot \exp \bigparen{- \frac{\delta^2}{3} \mu h^2}
    }
\end{observation}

This complicates the routing scheme. Any flow that doesn't finish in $2(1 + \delta)$ times the FCT of a single global flow should be backtracked and resent with new spraying decisions. However, this will only increase the maximum usable $h$:

\begin{table}[H]
\begin{tabular}{|l|l|l|l|l|}
\hline
$h$ & $\totalhosts$ & $p$ & $\delta$ & $3 \cdot 2(1 + \delta): 1$ \\ \hline
5 & $650$ & $0.992$ & $0.59$ & $3 \cdot 3.18 : 1$ \\ \hline
6 & $1332$ & $0.993$ & $0.731$ & $3 \cdot 3.461 : 1$ \\ \hline
7 & $2450$ & $0.992$ & $0.876$ & $3 \cdot 3.752 : 1$ \\ \hline
8 & $4160$ & $< 0$ & $1$ & bound is unusable for $h \geq 8$. \\ \hline
\end{tabular}
\end{table}

\begin{observation}
    We should consider Power-of-Two-Choices-like \cite{POTC} packet sprayers instead of simple oblivious ones for oversubscription bounds. Repeating for each group's $h^2$ hosts with $W$ packets each, if we randomly sample $h \geq d \geq 2$ outward queues on the leaf/spine layers and choose the least filled one, we can have the maximum load in an outward spine queue of:
    
    \eqalign {
        & P \bigparen{\max_{i \in \text{Group}} \bigparen{(\#0)_{i, \text{pe}(i)}} \leq \frac{W}{h^2} + \frac{\ln \ln h^2}{\ln d} + q - 1 + O(1)} \leq\\
        & \phantom{P \,\,\,} \leq 1 - \Theta(h^{-q})
    }
\end{observation}

For any integer $q \geq 1$. If we want all $|G|$ events (one for each group) to hold w.h.p., we need $q \geq 1 + 2$, as $|G| \in \Theta(h^2)$. Factorizing $W/h^2$, we can get an upper bound for $\alpha$ with probability $1 - \Theta(h^{-(q-2)})$:

\eqalign {
    & \alpha \leq \frac{W}{h^2} \bigparensq{1 + \underset{= \delta}{\frac{\ln \ln h^2 / \ln d + (q - 1) + O(1)}{W / h^2}}}
}

Which gives us an oversubscription bound of $2(1 + \delta) : 1$. In order to achieve a constant upper bound for $\delta$, we would need $\ln \ln h^2 / \ln d \geq \ln \ln h^2 / \ln h$ to decrease faster than $h^{-2}$, which is impossible.\\

In practice, the quality of the bound is highly dependent on the choice for $O(1)$ when $h$ is small. Because of it, the bound is generally too optimistic for a small $h$, and becomes increasingly too pessimistic as $h$ increases because of $h^{-2}$.\\

Several permutation pattern experiments from htsim follow. They specifically have these arguments:

\begin{table}[H]
    \begin{tabular}{|l|l|}
    \hline
    \# permutation traffic patterns & $100$ \\ \hline
    \# different run seeds per pattern & $10$ \\ \hline
    \# different global topologies for each Z matrix type (fig. \ref{oversubPermutationBars} only) & $5$ \\ \hline
    \end{tabular}
\end{table}

\begin{figure}[H]
    \centering
    \includegraphics[width=\linewidth]{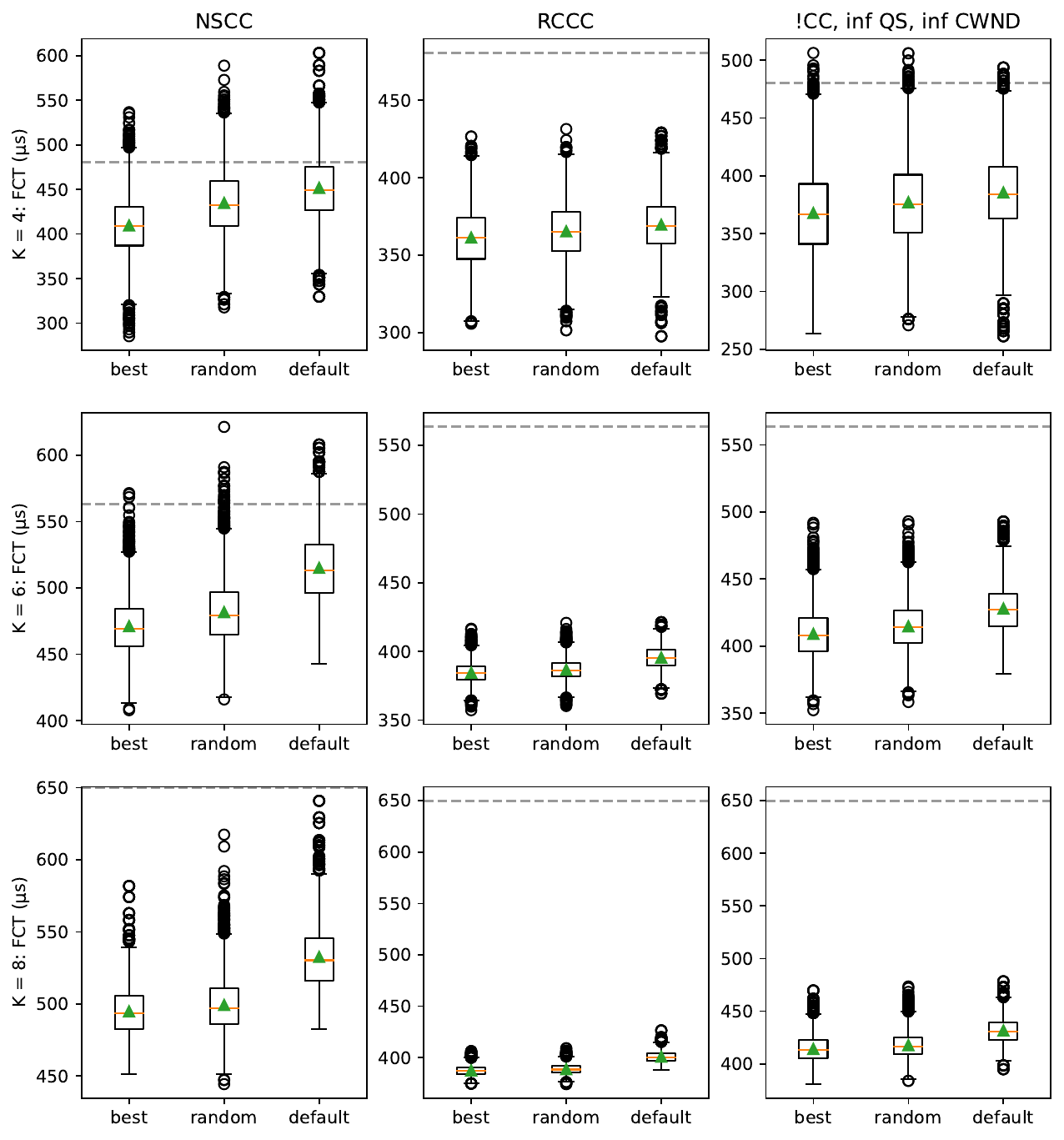}
    \caption{We investigate whether the Z matrix from {\bf notation \ref{notation_z}} can influence the performance of permutation patterns. The \texttt{best} global topologies have all $z_{\group(i), \group(\text{pe}(i))} = h-1$. We also bar plot \texttt{random} topologies, and the \texttt{default} configuration used in \cite{DragonflyPlus}.}
    \label{oversubPermutationBars}
\end{figure}

\begin{property}
    There are infinitely many half-radixes $h$ for which there exists a \texttt{best} topology with all $(z \circ \group)_{i, \text{pe}(i)} = h - 1$. Lemma 2.8 from \cite{FurinoNRB} proves that one exists if $h$ is a prime power.
\end{property}

\begin{figure}[H]
    \centering
    \includegraphics[width=\linewidth]{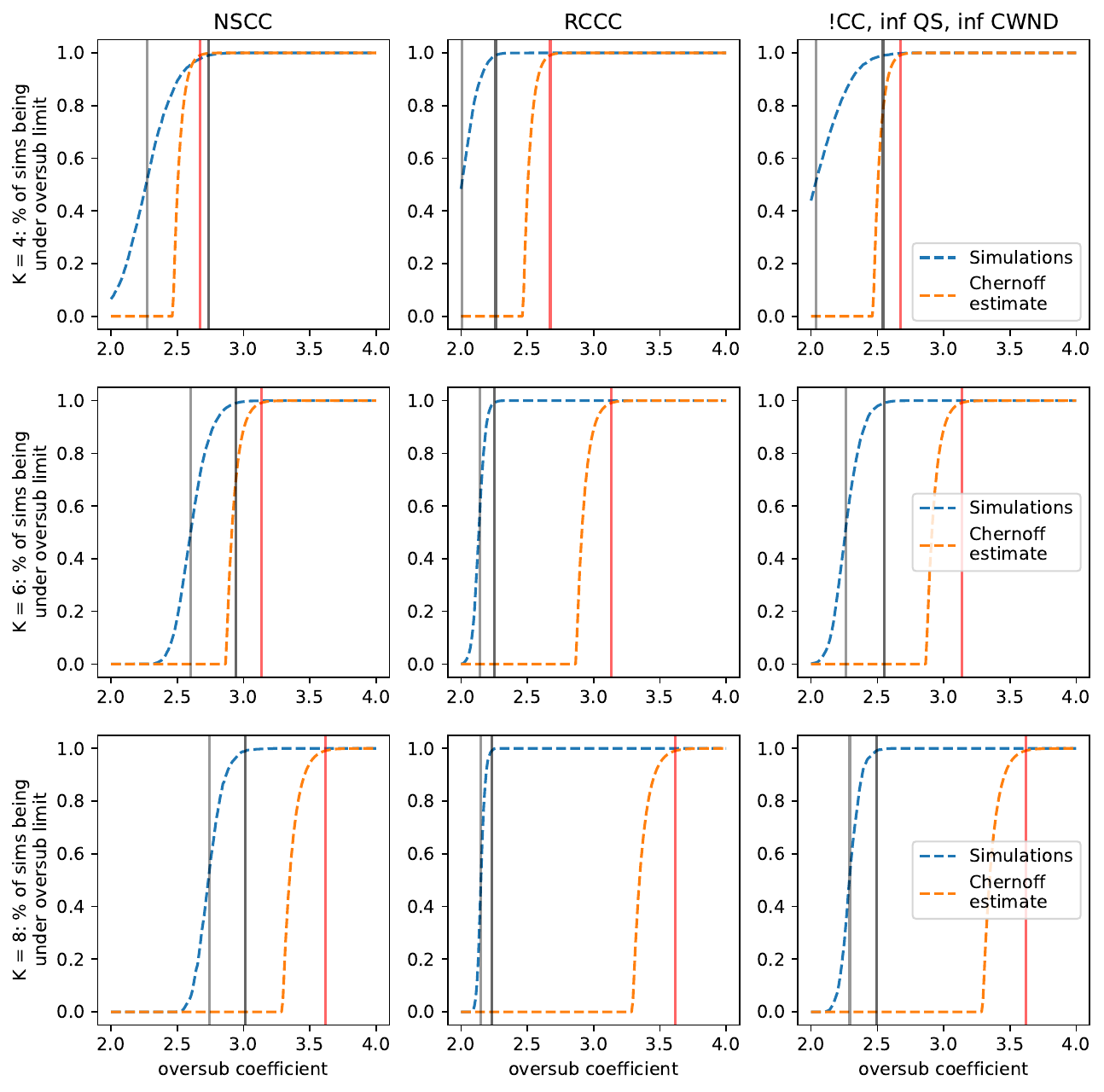}
    \caption{CDFs showing the fraction of permutation patterns whose oversubscription coefficient is under a certain value in $[2, 4]$. The blue curve represents results from htsim runs. The orange curve represents the Chernoff/Union Bound estimate from {\bf property \ref{property_chernoff_union_bound}} considering Valiant routing.}
    \label{oversubPermutationChernoff}
\end{figure}

The \texttt{default} configuration can be described as:

\begin{lstlisting}
if src_g < dst_g: # src_g, dst_g are group ids
    src_sw = (dst_g - 1) / h 
    dst_sw = src_g / h # src_sw, dst_sw are spine switch ids, local for each group.
\end{lstlisting}

The dashed horizontal line is the Chernoff estimate for the top $99\%$ from figure \ref{oversubPermutationChernoff}. There is a visible ranking between topology types for $k = 4$, but it dims between \texttt{best} and \texttt{random} for $k = 6$ and virtually disappears for $k = 8$.\\

The biggest speedup achieved against \texttt{default} is $\sim 7\%$, and occurs for $k = 8$ for sender-only CC. Note however that sender-only performs much worse for permutation patterns than receiver-only CC, where the speedup is unnotable.\\

The three rows represent the lowest three valid radixes $k \in \{4, 6, 8\}$. The columns represent different congestion control mechanisms being used, sender/reciever-only, or no CC with infinite queue size and CWND.\\

Each plot has an additional three vertical lines. The dim black line represents the lowest oversub coefficient which covers at least $50\%$ of runs. The darker black line counts $99\%$ of runs. The red line represents the Chernoff bound estimate for the top $99\%$.\\

The Chernoff bound should optimally be always more pessimistic than the actual runs, and should converge at $99\%$ at the same time. The $50\%$ bound should be hit by the runs at $2.0$ oversubscription (as it can be seen in figure \ref{oversubPermutationBars}, the median (orange line) and mean (green triangle) empirically overlap and can be used interchangeably here).\\

$W$ is always set up for the infinite queue size, even if the sender-only and receiver-only CC have the CWND set to $1$ and $1.5$ times the BDP respectively. While the Chernoff bound provides a great estimate for sender-only and infinite queue size when $k = 4$, it becomes visibly rough for all settings when $k$ increases. The receiver-only CC is closest to theory, reaching $50\%$ almost instantly after the $2.0$ coefficient.\\

Interestingly, the $50\%$ and $99\%$ lines always draw closer as $k$ increases, while the $99\%$ line moves slightly to the left in the receiver-only and no CC, infinite queue size settings. While we cannot prove that the $99\%$ will be reached earlier for higher values of $k$, we can empirically state that the $2.67: 1$ oversubscription bound for $k = 4$ holds with at least $99\%$ probability for any $k \geq 4$ in the receiver-only CC setting.

%% file: tex_parts/conclusion.tex
\section{Conclusion}\label{section_conclusion}

We have fixed and augmented the original Dragonfly+ \cite{DragonflyPlus} 2:1 oversubscription proof in expectation for the permutation traffic pattern, and in turn made it applicable to a vaster set of patterns. We only need all hosts to both send and receive at most line rate. This can be expanded to obtain an expected oversubscription estimate for any pattern. If any host sends at most $s$ times the line rate, and receives at most $r$ times the line rate, then the estimate is $2\max(s, r):1$. This works better if $s$ is closer to $r$, for example for all-to-all we have $s = r = \totalhosts - 1$ (host count).\\

We obtain reasonable high probability ($99\%$) oversubscription bounds for small values of $k$ on the permutation pattern: $2.67:1$ for $k = 4$ and $3.13:1$ for $k = 6$, when using sender-only CC. Interestingly, for receiver-only CC and no CC, infinite queue size and CWND, we empirically notice that $99\%$ of simulations finish with a smaller oversub bound, while their associated CDF rises faster for a higher radix $k$. If we can formally prove this, we could use the weak $2.67:1$ bound for $k = 4$ for any $k$.\\

Experimental results show that at least for a small $k$ and sender-only CC, the global topology can influence the oversubscription bound through the Z matrix, where $z_{i, j}$ counts how many intermediate groups have groups $i$, $j$ in the same spine. We observe speedups against the default global topology used in \cite{DragonflyPlus} of at most $\sim 7\%$ for $k \leq 8$. We hypothesize that for a large enough $k$ all possible global topologies will eventually perform similarly, at least in a theoretical infinite queue size and CWND setup.